\documentclass[a4paper,twoside,twocolumn]{article}
\pdfoutput=1

\newif\iftr 
\trtrue     
\newif\ifccs   
\ccstrue      

\usepackage{ifpdf}  
\ifpdf
    \usepackage[pdftex]{graphicx}
    \usepackage[colorlinks,linkcolor=blue,citecolor=blue,urlcolor=blue]{hyperref}
    \DeclareGraphicsExtensions{.pdf}
\else
    \usepackage{graphicx}
    \usepackage[hypertex]{hyperref}  
\fi

\usepackage{todonotes} 

\usepackage{
url       %
,fancyhdr 
,lastpage 
,enumerate 
,amsmath  
,amsfonts 
,amssymb  
, enumitem 
, array 
}
\usepackage[hang]{footmisc} 
\usepackage[noindent, arraystretch, fullpage]{setlengths} 
\usepackage{
own         
}

\graphicspath{{images/}}

\newcommand*{\metaauthori}{Bob Briscoe}

\newcommand*{\metashorttitle}{Rapid Signalling of Queue Dynamics}
\newcommand*{\metatitle}{{\large{Technical Report}}\\Rapid Signalling of Queue Dynamics}
\newcommand*{\metano}{TR-BB-2017-001}
\newcommand*{\metakeywords}{Data Communication, Networks, Internet, Control, Congestion Control, Quality of Service, Performance, Latency, Responsiveness, Dynamics, Algorithm, Active Queue Management, AQM, Signalling, Sojourn time, Queue delay, Service time, wait time, Expectation, Estimation, Explicit Congestion Notification, ECN}

\newcommand*{\metamaili}{\href{mailto:research@bobbriscoe.net}{research@bobbriscoe.net}}

\newcommand*{\metaaddress}{}

\newcommand*{\metaversion}{04}
\newcommand*{\metadate}{15 April 2019}

\hypersetup{                       
     pdfauthor = {\metaauthori%
     },
     pdftitle = {\metashorttitle},
     pdfsubject = {},
     pdfkeywords = {\metakeywords}
}%

\title{\metatitle}%
\author{\metaauthori%
\thanks{\metamaili, %
\metaaddress}%
}
\date{\metadate}%

\fancypagestyle{first}{%
\fancyhead[LO,RE]{}%
\fancyhead[LE,RO]{}%
\fancyhead[C]{}%
}%

\begin{document}
\bibliographystyle{alpha}%


\maketitle%
\thispagestyle{first}

\begin{abstract}
{\small\noindent%
Rather than directly considering the queuing delay of data, this memo focuses on reducing the delay that congestion signals experience within a queue management algorithm, which can be greater than the delay that the data itself experiences within the queue. Once the congestion signals are delayed, regulation of the load becomes more sloppy, and the queue tends to overshoot and undershoot more as a result, leading the data itself to experience greater peaks in queuing delay as well as intermittent under-utilization. 

Where the service rate of a queue varies, it is preferable to measure the queue in units of time not bytes. However, the size of the queued backlog can be measured in bytes and signalled at the last instant as data leaves the queue, whereas measuring queuing delay introduces inherent delay. This paper proposes 'scaled sojourn time', which scales queuing delay by the ratio between the backlogs at dequeue and enqueue. This is equivalent to scaling the sojourn time by the ratio of the arrival and departure rates averaged over the time spent in the queue. The paper also proposes the removal of delays due to randomness in the signal encoding.

}      
\end{abstract}
\ifccs{}%
%
%

\subsection*{CCS Concepts}
\textbf{\textbullet Networks} \(\to\) \textbf{Cross-layer protocols; Network algorithms; Network dynamics;}
\subsubsection*{Keywords}
\metakeywords
\fi{}%

\section{Introduction}\label{sigqdyntr_intro}

Much attention has been paid to reducing the delay experienced on the data path through packet networks. For instance, see sections II and IV of the extensive survey of latency reducing techniques in \cite{Briscoe14b:latency_survey}, which aim to reduce propagation delay, queuing delay, serialization delay, switching delay, medium acquisition delay and link error recovery delay.

Propagation and queuing delay are the largest contributors to the overall delay experienced by network data. Propagation delay can be reduced by structural techniques, such as server placement, but queuing delay is a result of subtle interactions due to the system design. 

Rather than directly considering the queuing delay of data, this memo focuses on reducing the delay that congestion signals experience within the queuing algorithm, which can be greater than the delay that the data itself experiences within the queue. Once the congestion signals are delayed, regulation of the load becomes more sloppy, and the queue tends to overshoot and undershoot more as a result, leading the data itself to experience greater peaks in queuing delay as well as intermittent under-utilization. Often peak delay is as critical as the average.

The focus here is on congestion signals transmitted from an active queue management (AQM) algorithm~\cite{Adams13:AQM_survey} using either drop or explicit congestion notification (ECN)~\cite{Floyd94:ECN}, which are the only standardized signalling protocols~\cite{IETF_RFC3168:ECN_IP_TCP} for end-to-end use over one of the two Internet protocols, IPv4 and IPv6.

These congestion signals experience delay consisting of the following elements:
\begin{itemize}[nosep]
	\item propagation delay (in common with the data)
	\item queuing delay (in common with the data)
	\item measurement delay: measuring the queue, as well as arrival and/or departure rates
	\item smoothing delay: filtering out fluctuations in measurements
	\item randomization delay: randomness is introduced to break up oscillations, but it requires longer to detect the underlying signal
	\item signal encoding delay: a number representing the signal is produced within an AQM algorithm but it then takes a longer time to transmit this number to the transport endpoints because it has to be compressed into one bit per packet using a unary encoding, otherwise the AQM would have to hold flow state
\end{itemize}

This memo focuses on reducing three of these: queuing, measurement and randomization. The other three are briefly surveyed in \S\,\ref{sec:related}.

The signal from an AQM can be subject to unnecessary queuing delay if it is applied during the enqueue process, so that it has to work its way through the queue before being transmitted to the line. In modern AQMs queuing delay is configured to be of the same order of magnitude as typical propagation delays. Therefore unnecessarily subjecting the congestion signal to the delay of the queue will add considerable sloppiness to the control loop.

Even if a signal is applied during the dequeue process, it can be based on a measurement that starts at the enqueue process. This measurement delay is inherent in the sojourn time technique that is becoming common for measuring the queue in modern AQMs. This memo proposes a simple technique to cut that measurement delay by using all the information available in the queue at the point a packet is dequeued. At the moment a packet is dequeued there is very little time for additional processing, so the technique is designed for minimal execution time.

The memo also proposes that randomization delay should be moved from the network to the end-system (just as smoothing delay has been similarly shifted in recent proposals (see \S\,\ref{sec:related})). This is a minor part of the memo that is orthogonal to the techniques to reduce the queuing and measurement aspects of signalling delays.

\section{Solutions}\label{sigqdyntr_solutions}

\subsection{Service Time of a Queue}\label{sec:svc_time}

In around 2012, it became recognized that one of the main problems with AQMs was the sensitivy of their  configuration to changing environments. For example:
\begin{itemize}
	\item access links often change their rate when modems retrain in response to interference. 
	\item a queue is part of a scheduling hierarchy and traffic in higher priority queues varies the capacity left for a lower priority queue, rapidly varying the drain rate that the AQM experiences.
	\item the capacity of radio links varies rapidly over time~\cite{McGregor10:Minstrel_TR}.
\end{itemize}

The CoDel algorithm~\cite{Nichols12:CoDel} proposed to solve this problem by measuring the the queue in units of time, rather than bytes. This made the configuration of the thresholds in the algoithm independent of the drain rate.

Actually, as far back as 2002, 
Kwon and Fahmy~\cite{Kwon02:Load_v_Queue_AQM} had advised that the queue should 
be measured in units of time. Also, in 2003, S{\aa}gfors \emph{et al} had 
modified the Packet Discard Prevention Counter (PDPC+~\cite{Sagfors03:PDPC_vary}) 
algorithm by converting queue length to queuing delay to cope with the varying 
link rate of 3G wireless networks.

PDCP still measured the queue in bytes, but then converted the result to time by dividing by the link rate, which it measured over a brief interval. 

CoDel proposed an elegant way to measure the service time of the queue by adding a timestamp to each packet's internal metadata on enqueue. Then at dequeue, it subtracted this timestamp from the system time. It called the result the sojourn time of the packet. It was pointed out that this sojourn time could be measured over an arbitrarily complex structure of queues, even across distributed input and output processors.

Because PIE~\cite{Pan13:PIE} was initially designed for implementation using existing hardware, it did not measure the service time of the queue directly using the time-stamping approach of CoDel.
Instead, like PDPC, it converted queue length to queuing delay using a regularly updated 
estimate of the link rate, measured over a set amount of packets. When there were insufficient packets in the queue to measure the link rate or the rate was varying rapidly, PIE's estimate of the link rate became stale. So in later specifications of PIE~\cite{Pan17:PIE}, it recommended the sojourn approach of CoDel that had been designed for software implementation.

Intially PIE also applied the congestion signal when it enqueued a packet. That is, it probabilistically dropped (or ECN-marked) the packet when it enqueued it. This signal then worked its way through the queue before being transmitted, which added another sojourn time before the signal reached the receiver, and subsequently the sender.

The queue length (in bytes or an equivalent unit), also called the backlog, can be measured instantaneously when a packet is enqueued or when it is dequeued. Whereas sojourn time can only be measured once a packet is dequeued. 

The matrix in \autoref{tab:added-delay} shows the delay added to the signal by various techniques for measuring queue delay (horizontal) and the two choices for where to apply the signal (vertical). It uses the following terminology: \(r\) is the duration used to sample the drain rate and \(s\) is the sojourn time. The right hand column shows the effective delay added by the simple estimation technique proposed in the following section.
\begin{table}[h]
\begin{center}
\begin{tabular}{m{0.17\columnwidth}|*{3}{m{0.2\columnwidth}}}
                     & \multicolumn{3}{c}{Technique to measure queue delay}\\
     where signal is applied  
                     & \(\frac{\mathrm{backlog}}{\mathrm{drain\_rate}}\)
                                                     & sojourn time
                                                                            &  scaled sojourn time\\\hline 
	at enq    & \(r/2 + s\)  & \(2s\)        & \(3s/2\)\\
	at deq    & \(r/2\)        & \(s\)           & \(s/2\)
\end{tabular}
\end{center}
\caption{Delay added to congestion signal by three different measurement techniques}%
\label{tab:added-delay}
\end{table}

The IETF specification of PIE~\cite{Pan17:PIE} recommends that the drain rate needs 16 packets to get a representative estimate, so \(r/2\) will be the serialization time of 8 packets, and it will become stale whenever there are less than 16 packets in the queue. In contrast, the sojourn time approaches apply down to a lone packet.

\subsubsection{Expected Service Time}\label{sec:inst_svc_time}

Whereas the amount of bytes in a queue can be measured at one any one instant, it takes one sojourn time to measure the sojourn time. Therefore, by the time the sojourn time has been measured it will be out of date, unless the arrival rate is the same as the departure rate, and the drain rate remains constant during the packet sojourn. 

The sojourn time measured when a packet reaches the head takes no account of any change in the queue while that packet is working towards the head. So, as a burst (or a reduction in drain rate) extends the queue, the sojourn time of the packet at the front of the burst (or the start of the reduction) will show no evidence of the queue that has built behind it. Sojourn time only fully measures the burst (or rate reduction) when the last packet of the burst reaches the head of the queue. 

Conversely, consider a queue that has been stable then the flow ends, so that no further packets arrive after a particular packet. Then, even when that packet was the last to leave from the head of the queue, its sojourn time would measure the stable queue delay when it arrived, because that would be how long it took to drain the queue. There would be no evidence of the now empty queue until traffic started again.

It is proposed to solve this problem by scaling the sojourn time by the ratio of the backlogs at dequeue and enqueue. That is, the expected service time at any instant will be:
\[\mathrm{E(svc\_time)} = \mathrm{sojourn\_time} \times \frac{\mathrm{(backlog\_deq)}}{\mathrm{(backlog\_enq)}},\]
where \(\mathrm{backlog\_at\_enq}\) can be written into the packet's metadata at enqueue.

\subsubsection{Rationale for Scaling Sojourn Time}\label{sec:inst_svc_time_justify}

As \autoref{tab:added-delay} shows, it takes time to measure a representative rate and it takes time to measure a time. Ideally, but perhaps na\"{\i}vely, just before forwarding a packet one could estimate the instantaneous drain rate as the serialization time of the previous head packet. Then one could calculate the instantaneous queuing delay as the instantaneous backlog divided by this instantaneous drain rate.

Then, for instance, if the arrival rate and drain rate have been constant while a packet works through the queue, but then the drain rate halves just before the packet is forwarded, the instantaneous queuing delay of the remaining queue will be double that measured by the sojourn time technique, whether or not it is scaled as proposed. 

However, there is no reason to believe that the latest instantaneous rate measurement is the best estimate of the rate at which the remaining queue will drain. For instance, a radio link is continually testing different rates to find which is the best and if a queue is continually yielding to a higher priority queue, it will proceed in fits and starts. 

Therefore, an estimate based on how the rate varied while the current head packet worked through the queue is not necessarily less correct than an estimate based on the drain rate at the instant the previous head packet departed.

Also, rather than having to arbitrarily choose a number of packets to measure over, sojourn time techniques automatically tune the most commonly used measurement duration to the most common queuing delay.

\begin{figure}[h]
	\centering
	\includegraphics[width=\columnwidth]{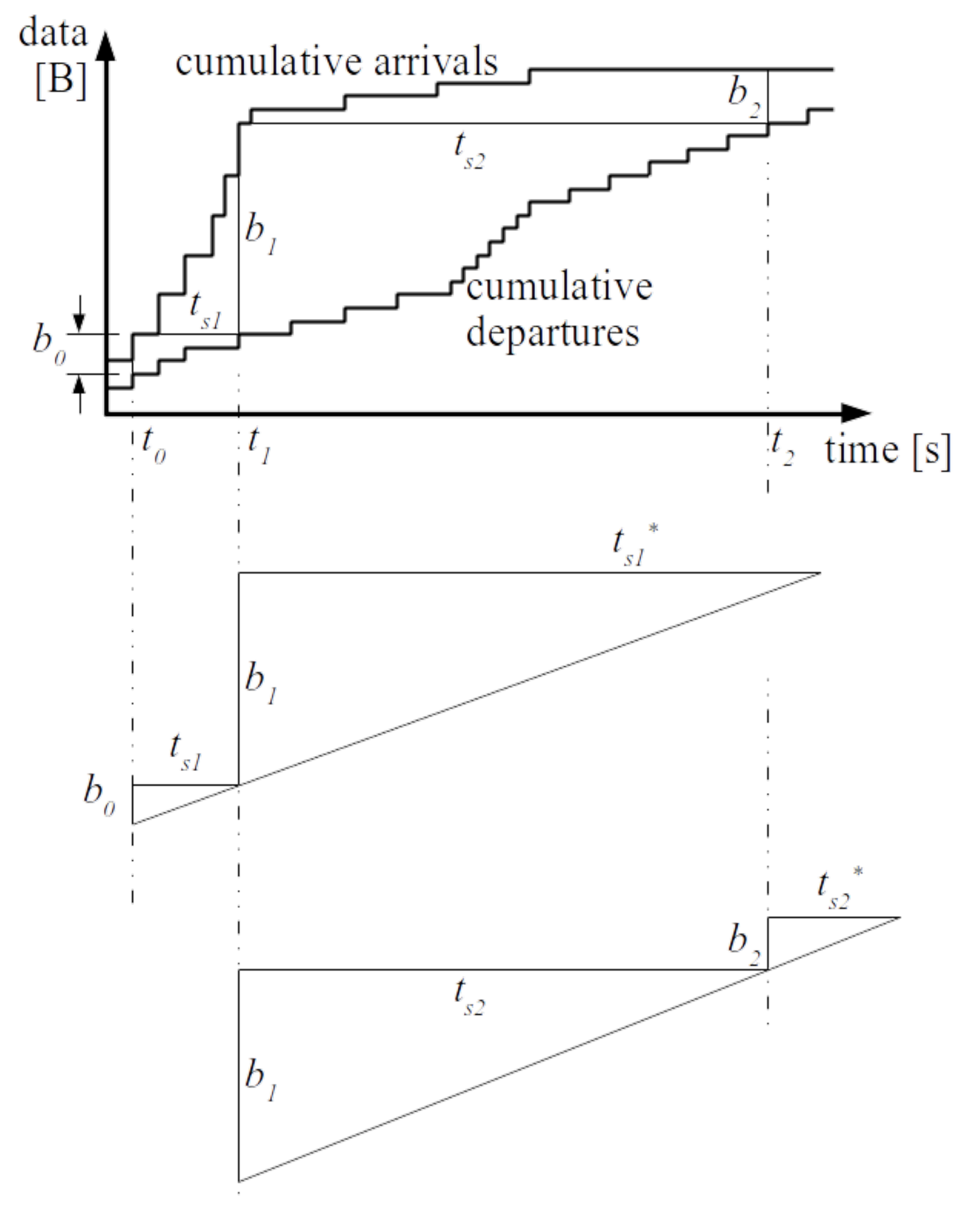}
	\caption{Rationale for Scaling Sojourn Time}\label{fig:scaled-sojourn}
\end{figure}

\autoref{fig:scaled-sojourn} visualizes the rationale for scaling the sojourn time. The two plots in the chart at the top of the figure show cumulative arrivals and departures of data in packets. Between times \(t_0\) and \(t_1\) a burst of packets arrives and between \(t_1\) and \(t_2\) a few packets arrive at first, then none. Over the whole time the departure rate is varying independently as, for example, a radio link would. At any time, for instance \(t_1\), the sojourn time  (\(t_{s1}\)) can be visualized as the horizontal distance back from the departures plot to the arrivals plot. And the backlog is shown as the vertical distance between the plots (\(b_1\)).

It can be seen that the sojourn time (\(t_{s1}\)) between \(t_0\) and \(t_1\) takes account of the departure rate, but not the arrival rate (the burst), during that time. It is proposed to scale the sojourn time by the ratio of the backlogs at departure and arrival of the packet. That is \(t_{s1}^* = t_{s1} b_1/b_0\). This scaled sojourn time uses all the latest information available at time \(t_1\). 

The schematic in the middle of the figure shows using similar triangles how scaled sojourn time is constructed.  The departure rate during the sojourn is represented by the slope of the smaller of the middle triangle. The larger triangle extrapolates that departure rate to predict the time (\(t_{s1}^*\)) that it will take for the most recent backlog to drain.

The lower schematic shows the situation at time \(t_2\). The actual sojourn time of the new head packet \(t_{s2}\) is slightly shorter than the prediction \(t_{s1}^*\). From this new actual sojourn time, a new prediction can now be constructed  from the slightly steeper rate slope. This time the backlog \(b_2\) has reduced  during the sojourn of the head packet, because there has been a lull in arrivals since \(t_1\). Therefore the formula predicts that the sojourn time will be scaled down relative to its measured value.

We will now return to time \(t_1\) and derive the scaled sojourn time algebraically, rather than geometrically. The departure rate during the sojourn of the head packet is
\begin{align}
	r_{d1} &= \frac{b_0}{t_{s1}}.\label{eqn:drate}
\intertext{The predicted sojourn time to drain the backlog at \(t_1\) is}
	t_{s1}^* &= \frac{b_1}{r_{d1}}.\notag
\intertext{Substituting from \autoref{eqn:drate}:}
				&= t_{s1} \frac{b_1}{b_0}.\label{eqn:sojourn}
\end{align}

Another way to think of the scaling is as the ratio of average arrival and departure rates during the sojourn, \(r_{a1}\) and \(r_{d1}\). The backlog at \(t_1\) can be expressed in terms of the arrival rate over the sojourn time:
\begin{align}
	b_1 &= t_{s1} r_{a1}.\label{eqn:backlog}
\intertext{Substituting this into \autoref{eqn:sojourn}, the scaled sojourn time at \(t_1\),}
	t_{s1}^* &= t_{s1} \frac{r_{a1}}{r_{d1}}
\end{align}
That is, scaling the sojourn time by the ratio between the backlogs at dequeue and enqueue is equivalent to scaling it by the ratio between the average arrival and departure rates between enqueue and dequeue.

\subsubsection{Implementing Scaled Sojourn Time}\label{sec:inst_svc_time_impl}

Some implementations choose not to do too much at dequeue, because there is limited time between the packet reaching the head of the queue and starting to be forwarded. Therefore, it could be challenging to measure the system time, subtract the stored timestamp then also scale the result by a ratio.

The following trick is likely to optimize execution of sojourn time scaling, although its efficiency will be machine-architecture-dependent:\\
{\small\texttt{qdelay <<= (lg(backlog\_deq) - lg(backlog\_enq)\\+ 1/2)}}\\
It is roughly equivalent to multiplying by the ratio between the backlogs, to the nearest integer power of 2.

The \texttt{<<=} operator bit-shifts \texttt{qdelay} to the left by the expression on the right. \texttt{lg()} is the logarithm function base 2. The expression bit-shifts \texttt{qdelay} to the left by the difference between the logs of the backlogs at enqueue and dequeue. The addition of 1/2 is necessary so that integer truncation of the result will round to the nearest integer, rather than always rounding down. 

The \texttt{clz()} function to count leading zeros could be used as a cheaper but more approximate equivalent, as follows:\\
{\small\texttt{qdelay <<= (clz(backlog\_enq) - clz(backlog\_deq))}}\\
This also avoids the need for any boundary checking code.

For example, if the \texttt{backlog\_*} variables are 32-bit unsigned integers and
\begin{itemize}[nosep]
	\item[] \texttt{backlog\_enq = 3000}, so \texttt{clz(3000)=20}
	\item[] \texttt{backlog\_deq = 30000}, so \texttt{clz(30000)=17}
\end{itemize}
Then
\begin{itemize}[nosep]
	\item[] \texttt{qdelay <<= 20 - 17}
\end{itemize}
is the same as
\begin{itemize}[nosep]
    \item[] \texttt{qdelay *= 2\^{}3},\\
\end{itemize}
which scales qdelay by 8, which approximates to \texttt{30,000/3,000 = 10} but is an integer power of 2. This is sufficient to scale the sojourn time to the correct binary order of magnitude, while still taking account of all the latest information in the queue.

However, \texttt{clz()} introduces truncation bias because it always rounds down, which could lead the result to be persistently out by up to \(\times2\) or \(/2\) for a particular target sojourn time. Using the \texttt{lg()}-based expression could be out by from \(\sqrt{2}\) to \(1/\sqrt{2}\), but with no bias---it is equally likely to be out either way.

A further rationale for scaling the sojourn time is that an implementation that is already measuring the sojourn time does not need any additional measurement code, because it already has to maintain a count of the backlog to do basic queue handling.

A high performance implementation will maintain the backlog of a queue by maintaining two variables (much like the two plots at the top of \autoref{fig:scaled-sojourn}):
\begin{itemize}[nosep]
	\item[] \texttt{count\_enq} written solely by the enqueue routine;
	\item[] \texttt{count\_deq} written solely by the dequeue routine
\end{itemize}	
Then the backlog can be measured as \texttt{count\_enq - count\_deq}. These two shared variables can be read from any routine, but they are only incremented by the routine that owns them, which avoids the performance hit of a mutual exclusion lock. The two counters monotonically increase like the system clock for the sojourn measurement, but at the rate of data transfer in and out respectively, not the rate of time passing. 

To implement scaling of the sojourn time, it is necessary to store \texttt{backlog\_enq} in the packet's metadata when the packet is enqueued. Then at dequeue it can be combined with \texttt{backlog\_deq} using the trick above.

\subsubsection{Distributed Queues}\label{sec:sojourn-distrib}

Using sojourn time leverages the advantage that it can be measured across a complex set of queues, including the case where the inital enqueue and the final dequeue routines are distributed across different machines or processors, as already mentioned.

This could include the case where the inputs are located on multiple client machines (e.g.\ mobile user equipment, WiFi stations, cable modems or passive optical network modems) while the output is a located at an aggregation node (e.g.\ a cellular base station (eNodeB)~\cite{Tan09:AQM_uplink_patent}, a WiFi access point (AP), a centralized controller for multiple WiFi APs, a cable modem terminal server (CMTS) or optical line termination (OLT) equipment), with a multiplexed access network between the clients and the aggregation node.

In this case, the timestamp and backlog at enqueue would have to be included in the protocol data units being transmitted between machines (e.g. within the L2 protocol), not just in packet metadata held within one machine's memory space. Also the aggregation node would need high priority (pref.\ non-blocking) access to the \texttt{count\_enq} variable on the input machine, in order to calculate \texttt{backlog\_deq}. Certain access network technologies, e.g.\ those for cellular radio access networks, already include such a control channel. The delay to access a control variable at the input machine from the output machine would be larger than that in a non-distributed system, but it would at least be a known, constant delay. So the control system would still provide robust metrics to control queuing in the data channel.

Of course, the sojourn time based on just a timestamp at enqueue could be written into PDUs to control any of the above distributed access networks, without the extra need for a non-blocking control channel. However, this would not provide the extra timeliness that scaled sojourn time would.

The measured sojourn time would include the delay before a packet or frame was given access to the shared medium, which would be the main cause of the backlog at the client queue. As well as the aggregation node using (scaled) sojourn time to apply congestion signalling within the final dequeue routine (effectively on behalf of the input queue), the aggregation node could also use (scaled) sojourn time to govern the scheduling algorithm for controlling each client's inward (upstream) access rate into the shared medium, by altering the rate at which it granted medium access slots to each client.

\subsubsection{Applicability of Scaled Sojourn Time}\label{sec:inst_svc_time_applic}

Scaling the sojourn time improves its timeliness, so it is applicable wherever sojourn time itself is useful.

It might be thought that an algorithm like the proportional integral (PI) controller\footnote{Used in QCN~\cite{IEEE802.1Qau:Ethernet_QCN}, PIE~\cite{Pan17:PIE}, PI2~\cite{DeSchepper16a:PI2} or the base AQM of DualPI2~\cite{Briscoe15e:DualQ-Coupled-AQM_ID}.} already takes account of the change in queuing delay between samples, so changing the queuing delay measurement itself seems redundant. However, scaling the sojourn time actually ensures that a PI algorithm takes account of the change between the latest queue delay measurements at each sample time, not between two outdated measurements.

It might also be thought that PI controllers do not need to care so much about instantaneous measurements, because they are maintaining the fairly large queue that is needed by classic TCP algorithms like Reno, Cubic, Compound or BBR. However, even though a PI algorithm only samples the queue fairly infrequently (relative to packet serialization time), using an out of date queue metric makes it necessary to introduce extra heuristic code to deal with the resulting sloppiness.

For instance, in the case of PIE~\cite{Pan17:PIE}, some heuristic code suppresses any drop once the last sample of queue delay falls below half the target delay.\footnote{As long as some other conditions hold that are not important here.} This is an attempt to suppress drop when the queue is draining after the load has gone idle. However, it is ineffective if sojourn time is used to measure the queue, because the sojourn time does not reduce until after the last packet (as explained earlier). Scaling the sojourn time whenever it is sampled should eliminate the need for this metric because it takes account of the reducing backlog as the queue drains. Indeed, this was the original motivation for developing the scaled sojourn time metric.

Scaling the sojourn time is also highly applicable to the CoDel algorithm for the same reasons---sojourn time fails to take account of the evolution of the queue after the head packet was enqueued. In CoDel's case, sojourn time is measured per packet, so the scaling would have to be applied per packet. Nonetheless, the trick above at least minimizes the cycles required.

Scaling the sojourn time should also be applicable to a simple low threshold algorithm like the time-based threshold recommended for DCTCP in~\cite{Bai16:MQ-ECN} and proposed as the native AQM for more general, so-called `L4S' traffic in DualPI2~\cite{Briscoe15e:DualQ-Coupled-AQM_ID}, where L4S stands for Low Latency, Low Loss, Scalable throughput. It would be applicable whether the threshold is a simple step, or a probabilistic ramp like the RED function (but based on instantaneous sojourn time, not smoothed queue length), or a deterministic ramp or convex function of instantaneous queueing delay. However, given these schemes are intended to keep queue delay very low, there is less scope for widely varying queue dynamics, so the cost of the extra processing might not prove to be worth the benefit.

Scaling the sojourn time of a queue applies to many types of queue, not just packet queues, as long as the size of each job is quantified in common units that are additive. Examples include, but are not limited to, queues of datagrams, frames or packets, as well as message queues, call-server queues, computer process scheduling queues, storage queues (e.g. SSD or disk), workflow queues for mechanical or human-operated stages of tasks. 

As well as dropping or ECN-marking, different sanctions could be applied using the same basic ideas. Examples include, but are not limited to: truncating or otherwise damaging the data or checksum of a message or packet but preserving the information necessary for delivery; rerouting; delaying; downgrading the class of service; and tagging.

\subsection{Removing Randomness Delays}\label{sec:rand_delay}

One of the main motivations for the design of Random Early Detection (RED)~\cite{Floyd93:RED} was to introduce randomness to break up synchronization between the sawteeth of TCP flows driving the same queue. This still remains an important requirement for all AQM algorithms~\cite{Baker15:AQM_Recommendations}.

With clean-slate approaches such as DCTCP in private networks, or incrementally deployable clean-slate approaches like L4S~\cite{Briscoe16a:l4s-arch_ID} for the public Internet, requirements for the network and for end-systems are still in the process of definition. In these clean-slate or slightly dirty clean-slate cases, it would be possible to require the sender's congestion control to dither its response to congestion signals, so that it would not be necessary to introduce randomness in the network, which adds uncertainty and therefore delay to the congestion signalling channel. 

Any AQM that probabilistically signals congestion with probability \(p\) could deterministically signal congestion by introducing an interval of \(1/p\) packets between each drop or mark. 

The determinism would be lost wherever the AQM was controlling flows mulitplexed within one queue without per-flow state, because assignment of each deterministic congestion signal to each flow would become randomized by even slightly random packet arrivals from the different flows~\cite{Briscoe15d:PIE_rvw}.

Nonetheless, whenever a flow is on its own in an AQM, which is a common case for the traffic patterns in many access network designs,  deterministic congeston signalling would reduce signalling delay. This could particularly ease the design of new flow-start algorithms, where the flow introduces microbursts or chirps to sense at what level it starts to congest the link.

\section{Related Work}\label{sec:related}

Scaling sojourn time seems superficially similar to combined enqueue and dequeue ECN marking (CEDM)~\cite{Shan17:CEDM}, because CEDM marks a packet at enqueue if the queue is over a threshold, but then unmarks it at dequeue if the backlog has dropped below the threshold. However the two are significantly different. Firstly, CEDM has to be based on queue length in order to mark at enqueue. But also CEDM is intentionally asymmetric, in that it unmarks packets if the backlog at dequeue has dropped below the threshold, but it does not mark packets at dequeue if they have risen above the threshold. In contrast, scaling sojourn time is deliberately symmetric, meaning it compensates for growth or shrinkage of the backlog (\autoref{fig:scaled-sojourn}).

PDPC+~\cite{Sagfors03:PDPC_vary} and CoDel~\cite{Nichols12:CoDel}, which is very similar, use a deterministic rather the probabilistic algorithm to encode the congestion signal. However, they do not propose a way to introduce randomness in the end-systems instead. Therefore, they are likely to be prone to synchronization effects.

The introduction enumerated six causes of delay to congestion signals and highlighted three that this memo would focus on. The other sources of signalling delay have been considered in other work which is briefly surveyed below.

\paragraph{Propagation Delay:} Numerous proposals have been made to speed up signalling by sending the signal from the queue back against the flow of traffic, direct to the sender. This can be done in a pure L2 network, e.g. backwards congestion notification (BCN) in IEEE 802.1Qau~\cite{IEEE802.1Qau:Ethernet_QCN} a.k.a.\ Quantized Congestion Notification (QCN), which is now rolled into 802.1Q-2011 and 802.1Q-2014. However, in general signalling backwards is problematic in IP networks, amongst other reasons because the sender has to accept out-of-band packets from any arbitrary source in the middle of the network, which makes it vulnerable to DoS attacks~\cite{IETF_RFC6633:ICMP_SQ_Depr}. 

Therefore, here we will assume that signals are piggy-backed on the forward traffic flow then fed back to the sender via the receiver. However, this does not preclude a solution to the problems of backwards congestion notification.

\paragraph{Smoothing Delay:} AQMs designed for the Internet's classic congestion controls (TCP Reno, Cubic, Compound, etc.) filter out fluctuations in the queue by smoothing it before using the smoothed measurement as a measure of load to drive the congestion signal. DCTCP proposed to smooth the signal at the sender, so that the network could send out the signal immediately, without smoothing. This allows the sender to receive the signal without smoothing delay, which is particularly useful in cases where the sender might not need to smooth the signal itself, e.g.\ to detect overshoot when accelerating to start a new flow. Shifting the smoothing function from the network to the sender also makes sense because the network does not know the round trip time (RTT) of each flow, so it has to smooth over the maximum likely RTT. Whereas a sender knows its own RTT and can smooth over this timescale.

Here, we will assume no smoothing delay in the network, but that is orthogonal to the approaches proposed, which do not preclude network-based smoothing.

\paragraph{Signal encoding delay:} Previous research has proposed to change the IP wire protocol to provide more bits to signal congestion. Nonetheless, it has been pointed out that the delay of a unary encoding is inversely proportional to the value being encoded, and the congestion window of a scalable congestion control is also inversely proportional to the value of the congestion signal. So, as flow rates (and consequently congestion windows) increase over the years, at least in general the delay to encode the signal does not increase.

Therefore, here we assume a standard unary encoding of congestion signals. This does not preclude other encodings, e.g. the multi-bit encoding of QCN or minor alterations to the decoding to avoid saturation, such as that in \cite{Briscoe17a:CC_Tensions_TR}.

\section{Further Work}

These ideas might not be novel, but no concerted effort has been made to search the literature. The ideas have not been evaluated either.

\section{Acknowledgements}\label{sigqdyntr_acks}

The scaling of the service time of the queue is based on discussions with Henrik Steen, an MSc student of the author, in Nov 2016 \& May 2017.      

\addcontentsline{toc}{section}{References}

{\footnotesize%
\bibliography{aqm-details}}

\newcommand{\etalchar}[1]{$^{#1}$}
\begin{thebibliography}{DSBEBT17}

\bibitem[Ada13]{Adams13:AQM_survey}
Richelle Adams.
\newblock {Active Queue Management: A Survey}{}.
\newblock {\em IEEE Communications Surveys \& Tutorials}, 15(3):1425--1476,
  2013.

\bibitem[BBP{\etalchar{+}}16]{Briscoe14b:latency_survey}
Bob Briscoe, Anna Brunstrom, Andreas Petlund, David Hayes, David Ros, Ing-Jyh
  Tsang, Stein Gjessing, Gorry Fairhurst, Carsten Griwodz, and Michael Welzl.
\newblock {Reducing Internet Latency: A Survey of Techniques and their
  Merits}{}.
\newblock {\em IEEE Communications Surveys \& Tutorials}, 18(3):2149--2196, Q3
  2016.
\newblock (publication mistakenly delayed since Dec 2014).

\bibitem[BCCW16]{Bai16:MQ-ECN}
Wei Bai, Li~Chen, Kai Chen, and Haitao WuHaitao.
\newblock {Enabling ECN in Multi-Service Multi-Queue Data Centers}{}.
\newblock In {\em 13th USENIX Symposium on Networked Systems Design and
  Implementation (NSDI 16)}, pages 537--549, Santa Clara, CA, March 2016.
  USENIX Association.

\bibitem[BDS17]{Briscoe17a:CC_Tensions_TR}
Bob Briscoe and Koen De~Schepper.
\newblock {Resolving Tensions between Congestion Control Scaling
  Requirements}{}.
\newblock Technical Report TR-CS-2016-001, Simula, July 2017.

\bibitem[BEDSB17]{Briscoe16a:l4s-arch_ID}
Bob Briscoe~(Ed.), Koen De~Schepper, and Marcelo Bagnulo.
\newblock {Low Latency, Low Loss, Scalable Throughput (L4S) Internet Service:
  Architecture}{}.
\newblock Internet Draft draft-ietf-tsvwg-l4s-arch-00, Internet Engineering
  Task Force, May 2017.
\newblock (Work in Progress).

\bibitem[BF15]{Baker15:AQM_Recommendations}
Fred Baker and Gorry Fairhurst.
\newblock {IETF Recommendations Regarding Active Queue Management}{}.
\newblock Request for Comments RFC7567, RFC Editor, July 2015.

\bibitem[Bri15]{Briscoe15d:PIE_rvw}
Bob Briscoe.
\newblock {Review: Proportional Integral controller Enhanced (PIE) Active Queue
  Management (AQM)}{}.
\newblock Technical Report TR-TUB8-2015-001, BT, May 2015.

\bibitem[DSBEBT17]{Briscoe15e:DualQ-Coupled-AQM_ID}
Koen De~Schepper, Bob Briscoe~(Ed.), Olga Bondarenko, and Ing-Jyh Tsang.
\newblock {DualQ Coupled AQM for Low Latency, Low Loss and Scalable
  Throughput}{}.
\newblock Internet Draft draft-ietf-tsvwg-aqm-dualq-coupled-01, Internet
  Engineering Task Force, July 2017.
\newblock (Work in Progress).

\bibitem[DSBTB16]{DeSchepper16a:PI2}
Koen De~Schepper, Olga Bondarenko, Ing-Jyh Tsang, and Bob Briscoe.
\newblock {PI\(^2\) : A Linearized AQM for both Classic and Scalable TCP}{}.
\newblock In {\em Proc. ACM CoNEXT 2016}, pages 105--119, New York, NY, USA,
  December 2016. ACM.

\bibitem[FE10]{IEEE802.1Qau:Ethernet_QCN}
Norm Finn~(Ed.).
\newblock {IEEE Standard for Local and Metropolitan Area Networks---Virtual
  Bridged Local Area Networks - Amendment: 10: Congestion Notification}.
\newblock Draft standard 802.1Qau, IEEE, April 2010.

\bibitem[FJ93]{Floyd93:RED}
Sally Floyd and Van Jacobson.
\newblock {Random Early Detection Gateways for Congestion Avoidance}{}.
\newblock {\em IEEE/ACM Transactions on Networking}, 1(4):397--413, August
  1993.

\bibitem[Flo94]{Floyd94:ECN}
Sally Floyd.
\newblock {TCP and Explicit Congestion Notification}{}.
\newblock {\em ACM SIGCOMM Computer Communication Review}, 24(5):10--23,
  October 1994.
\newblock (This issue of CCR incorrectly has '1995' on the cover).

\bibitem[Gon12]{IETF_RFC6633:ICMP_SQ_Depr}
Fernando Gont.
\newblock {Deprecation of ICMP Source Quench Messages}{}.
\newblock Request for Comments 6633, RFC Editor, May 2012.

\bibitem[KF02]{Kwon02:Load_v_Queue_AQM}
Minseok Kwon and Sonia Fahmy.
\newblock {A Comparison of Load-based and Queue-based Active Queue Management
  Algorithms}{}.
\newblock In {\em Proc. Int'l Soc. for Optical Engineering (SPIE)}, volume
  4866, pages 35--46, 2002.

\bibitem[MS10]{McGregor10:Minstrel_TR}
Andrew McGregor and Derek Smithie.
\newblock {Rate Adaptation for 802.11 Wireless Networks: Minstrel}{}.
\newblock http://blog.cerowrt.org/papers/minstrel-sigcomm-final.pdf, 2010?
\newblock (Rejected conference submission).

\bibitem[NJ12]{Nichols12:CoDel}
Kathleen Nichols and Van Jacobson.
\newblock {Controlling Queue Delay}{}.
\newblock {\em ACM Queue}, 10(5), May 2012.

\bibitem[PNB{\etalchar{+}}17]{Pan17:PIE}
Rong Pan, Preethi Natarajan, Fred Baker, Greg White, Bill Ver~Steeg, Mythili
  Prabhu, Chiara Piglione, and Vijay Subramanian.
\newblock {PIE: A Lightweight Control Scheme To Address the Bufferbloat
  Problem}{}.
\newblock Request for Comments RFC 8033, RFC Editor, February 2017.

\bibitem[PPP{\etalchar{+}}13]{Pan13:PIE}
Rong Pan, Preethi Natarajan~Chiara Piglione, Mythili Prabhu, Vijay Subramanian,
  Fred Baker, and Bill Ver~Steeg.
\newblock {PIE: A Lightweight Control Scheme To Address the Bufferbloat
  Problem}{}.
\newblock In {\em High Performance Switching and Routing (HPSR'13)}. IEEE,
  2013.

\bibitem[RFB01]{IETF_RFC3168:ECN_IP_TCP}
K.~K. Ramakrishnan, Sally Floyd, and David Black.
\newblock {The Addition of Explicit Congestion Notification (ECN) to IP}{}.
\newblock Request for Comments 3168, RFC Editor, September 2001.

\bibitem[SLMP03]{Sagfors03:PDPC_vary}
M.~S{\aa}gfors, R.~Ludwig, M.~Meyer, and J.~Peisa.
\newblock {Buffer Management for Rate-Varying 3G Wireless Links Supporting TCP
  Traffic}{}.
\newblock In {\em Proc Vehicular Technology Conference}, April 2003.

\bibitem[SR17]{Shan17:CEDM}
Danfeng Shan and Fengyuan Ren.
\newblock {Improving ECN Marking Scheme with Micro-burst Traffic in Data Center
  Networks}{}.
\newblock In {\em Proc. IEEE Conference on Computer Communications
  (Infocom'17)}, May 2017.

\bibitem[TST10]{Tan09:AQM_uplink_patent}
Yifeng Tan, Riikka Susitaival, and Johan Torsner.
\newblock {Active Queue Management for Wireless Communication Network
  Uplink}{}.
\newblock Patent WO2010107355, 2010.

\end{thebibliography}


\onecolumn%
\addcontentsline{toc}{part}{Document history}
\section*{Document history}

\begin{tabular}{|c|c|c|p{3.5in}|}
 \hline
Version &Date &Author &Details of change \\
 \hline\hline
00A                   & 04 Sep 2017 &Bob Briscoe &First draft.\\\hline%
01                   & 05 Sep 2017 &Bob Briscoe &First complete version.\\\hline%
02                  &07 Sep 2017     &Bob Briscoe &Added a couple of refs. Qualified claims about clz().\\\hline%
03                  &16 Jan 2018     &Bob Briscoe &Completed the algebraic rationale for scaling sojourn time.\\\hline%
\metaversion &\metadate     &Bob Briscoe &Added abstract\\\hline%
\hline%
\end{tabular}

\end{document}